\documentclass[12pt]{article}
\usepackage{graphicx}
\usepackage{amssymb,mathtools}
\usepackage{amsmath}
\usepackage{color}
\usepackage{tabularx}
\usepackage{enumitem}
\usepackage{natbib}
\usepackage{hyperref}
\usepackage{dsfont}
\usepackage{algorithm}
\usepackage{algpseudocode}
\usepackage{multirow}
\usepackage{threeparttable}
\usepackage{lscape}
\usepackage{booktabs}
\usepackage{url}
\usepackage{caption2}

\newcommand{\blind}{0}



\newcommand{\abs}[1]{\left\lvert#1\right\rvert}
\newcommand{\norm}[1]{\left\lVert#1\right\rVert}

\newtheorem{thm}{Theorem}

\newtheorem{assum}{Assumption}

\begin{document}

\def\spacingset#1{\renewcommand{\baselinestretch}%
{#1}\small\normalsize} \spacingset{1}


\if0\blind
{
  \title{\bf Spatial Interference Detection in Treatment Effect Model}
  \author{Wei Zhang\\
    School of Mathematical Sciences, Center for Statistical Science, \\Peking University, Beijing, China\\
    Ying Yang \\
    Academy of Mathematics and Systems Science, \\ 
    Chinese Academy of Sciences, Beijing, China\\
    Fang Yao\thanks{The first two authors contribute equally. Corresponding author: \href{fyao@math.pku.edu.cn}{fyao@math.pku.edu.cn}.\hspace{.2cm}} \\
    School of Mathematical Sciences, Center for Statistical Science, \\Peking University, Beijing, China}
  \maketitle
} \fi

\if1\blind
{
  \bigskip
  \bigskip
  \bigskip
  \begin{center}
    {\LARGE\bf Spatial Interference Detection in Treatment Effect Model}
\end{center}
  \medskip
} \fi

\bigskip
\begin{abstract}
Modeling the interference effect is an important issue in the field of causal inference. Existing studies rely on explicit and often homogeneous assumptions regarding interference structures. In this paper, we introduce a low-rank and sparse treatment effect model that leverages data-driven techniques to identify the locations of interference effects. A profiling algorithm is proposed to estimate the model coefficients, and based on these estimates, global test and local detection methods are established to detect the existence of interference and the interference neighbor locations for each unit. We derive  the non-asymptotic bound of the estimation error, and establish theoretical guarantees for the global test and the accuracy of the detection method in terms of Jaccard index. Simulations and real data examples are provided to demonstrate the usefulness of the proposed method.
\end{abstract}

\noindent%
{\it Keywords:} interference detection, high dimension, low-rank and sparse model.
\vfill

\newpage
\spacingset{1.5} 

\section{Introduction}
\label{sec:intro}

In the classical potential outcome framework for causal inference, a basic assumption is that there is no interference between individuals, which is the key of the stable unit treatment value assumption \citep[SUTVA]{RN101}. However, this assumption is often violated in practical scenarios, such as in vaccine trials \citep{RN106,RN107}, air quality studies \citep{RN117,RN16} and policy evaluation problems \citep{shi2022multi,luo2024policy}. To ensure the validity of inference results, it becomes imperative to assess the presence of interference and, more critically, to comprehend the structure of these spillover effects. With this motivation, we establish the framework of spatial interference detection in this work.

As demonstrated by  \cite{RN31} and \cite{eckles2017design}, ignoring interference effects (IE) can lead to biased inference results. There are a line of works studying the test of interference existence. \cite{Rosenbaum2007} formulated two sharp null hypotheses that imply that the stable unit treatment value assumption does not hold. \cite{Aronow2012A} and \cite{Athey2018} explicitly tackled testing for the nonsharp null hypothesis that the stable unit treatment value assumption holds. \cite{2019Testing} proposed a hierarchical randomized experimental design to test the presence of arbitrarily structured interference. However, these works can not detect the specific locations of interference neighbors for each unit (unit $j$ is considered an interference neighbor of unit $k$ if the treatment on unit $j$ affects the outcome of unit $k$).

 To accurately evaluate treatment effects, existing studies assume that the interference neighbor set of each unit is known. 
The common types of interference structure assumptions can be categorized into three groups.
The first category is the partial interference, which assumes that individuals can be divided into different clusters, and the interference effect only exists within each cluster \citep{RN105,RN106,RN123,RN107,RN124,RN127}. The second category is the local network interference, which assumes that the interference effect exists in the local network surrounding each node \citep{RN31,RN17,RN130,RN129}.
The third type is to combine the interference structure with the underlying mechanism. For example, \cite{RN12}, \cite{RN11} and \cite{RN13} characterized interference effects using the market equilibrium; \cite{RN113} assumed that the interference effect in time series problems has a $d$-order lag structure.
Nevertheless, these assumptions may have limitations in the assumed spatial homogeneity for the interference structures, whereas in reality, different individuals may possess distinct interference structures.

In this paper, we propose a spatial interference detection method that can identify the interference structure for each unit, serving as the preliminary step for studying treatment effects. Suppose that the area of interest can be divided into $R$ rows and $C$ columns of grids. Each grid serves as a unit and has $n$ observations. The modeling of IE for a unit is naturally a high-dimensional problem. We assume that IE possesses a sparse structure, i.e., the number of interference neighbors of a unit is significantly smaller than the total number of units. 
It is noteworthy that the aforementioned interference  assumptions can  be regarded as specific forms of the sparse interference structure.
We refer to the effect of one unit's treatment on  itself as the direct effect (DE). Noting that the spatial structure of DE is usually determined by several latent-variables in practice \citep{chernozhukov2021inference}, we thus assume that DE is determined by a low-rank structure over space. Then we propose to use a profiling algorithm to estimate the model coefficients. Based on these estimates, we can conduct a global test to detect the existence of IE. Consequently high-dimensional signal detection methods such as  the binary and re-search \citep[BiRS]{zhang2023binary} and stepdown \citep{Romano2005} can be applied to detect the locations of neighbors for each unit.


The contributions are summarized as follows. Firstly, this work pioneers the detection problem of interference neighbor locations for each unit, which helps avoid mis-specification of interference structures and lays the foundation for subsequent analysis. In particular, the sparse assumption efficiently reduces the model dimension and based on the proposed method, we can conduct a post-detection estimation of average treatment effects (ATE), as discussed in Section \ref{sec:detection}. Secondly, we expand the high-dimensional mean signal detection to the sparse regression setting. Different from the mean signal detection problem, sparse regression entails optimization with tuning parameters, rendering the associated inference more challenging. We propose  a conditional bootstrap method to facilitate inference, based on the theoretical properties of the sparse estimate. Thirdly, we establish solid theoretical guarantees for the proposed method. We derive the error bounds of the estimates and establish the inference validation including size and power analysis. Under mild assumptions, the proposed method can consistently detect the signal locations of interference neighbors for all units. Lastly, the proposed method addresses three main challenges for practical applications: (i) the model allows for interference with heterogeneous forms, making it applicable to spatial causal inference problems with, e.g., geographical differences and/or economic imbalances; (ii) the low-rank and sparse constraints of the model make it suitable for scenarios with small effect sizes; (iii) detecting interference structure unit-wisely is instrumental in understanding the mechanisms through which the treatment surpasses the control, thereby paving the way for the  development of more effective strategies, particularly when budget constraints exist. This includes targeting treatment towards units with large positive effects, maximizing the impact within limited resources. We evaluate the performance of the proposed test and detection by using extensive simulated and real data experiments, which offers empirical support for our findings.

The rest of the paper is organized as follows. Section \ref{sec:model} introduces the low-rank and sparse treatment effect model and outlines the estimation procedure. We also present the methods for global testing and IE location detection. In Section \ref{sec:thm}, we derive the error bounds of the estimates, and perform size and power analysis for the global test. We then describe the analysis of detection accuracy for the stepdown and BiRS algorithms. Comprehensive simulation studies based on the ChinaHighAirPollutants data are conducted in Section \ref{sec:simulation}, and two real data examples for IE location detection are presented in Section \ref{sec:applications}. Concluding remarks are provided in Section \ref{sec:conclude}, and technical proofs of main theorems are deferred to Supplementary Material.

\section{Low-rank and Sparse Treatment Effect Model}

Large-scale environmental and epidemiological studies often use spatially referenced data to examine the effect of treatments. Suppose that the experimental area is partitioned into $R$ rows and $C$ columns of distinct, non-overlapping units. Let $M_{rc}$ be a binary variable representing the treatment assignment to unit $(r,c)$ and $Y_{rc}$ the observed outcome of unit $(r,c)$. Let $M$ and $Y$ be the corresponding $RC$-dim vectors in the whole experimental area. When SUTVA does not hold, $Y_{rc}$ depends on $M$ instead of $M_{rc}$. 

\subsection{Potential outcome framework and interference effect}

We now use the potential outcome notations to introduce the interference effect. Let $Y_{rc}(M)$ be the potential outcome for unit $(r,c)$ under treatment $M$.
As pointed out by \cite{rubin1986comment}, the potential outcome is well defined only if the following assumption holds:

\begin{assum}
(Consistency Assumption, CA) For each unit $(r,c)$, the observed outcome satisfies $Y_{rc}=Y_{rc}(M)$.
\end{assum}

Let $M_{-rc}$ be the vector of treatments for all units other than $(r,c)$.
The individual treatment effect (ITE) for unit $(r,c)$ is defined as 
\begin{equation*}
\text{ITE}_{rc}=\mathbb{E}\left\{Y_{rc}(M_{rc}=1,M_{-rc}=1)-Y_{rc}(M_{rc}=0,M_{-rc}=0)\right\}.
\end{equation*}
This expression can be decomposed as follows:
\begin{equation*}
\text{ITE}_{rc} = \underbrace{\mathbb{E}\left\{Y_{rc}(1,1)-Y_{rc}(0,1)\right\}}_{\text{DE}_{rc}} + \underbrace{\mathbb{E}\left\{Y_{rc}(0,1)-Y_{rc}(0,0)\right\}}_{\text{IE}_{rc}},
\end{equation*}
where $\text{DE}_{rc}$ represents the direct effect of $M_{rc}$ on the outcome of unit $(r,c)$ and $\text{IE}_{rc}$ captures the interference effect of $M_{-rc}$ on $Y_{rc}$.
The population level effects are then given by 
$$\text{ATE}=\frac{1}{RC}\sum_{r,c}\text{ITE}_{rc},\ \text{DE}=\frac{1}{RC}\sum_{r,c}\text{DE}_{rc} \ \text{and} \ \text{IE}=\frac{1}{RC}\sum_{r,c}\text{IE}_{rc}.$$

In this paper, we focus on detecting the heterogeneous interference structure. For any unit $(r,c)$, let $\mathcal{N}_{rc}\subset\{(r',c')\neq (r,c):1\le r'\le R, 1\le c' \le C\}$ be an index subset dependent on $(r,c)$, and let $\mathcal{N}_{-rc}$ denote all units not in $\{i\}\cup\mathcal{N}_{rc}$. 
We impose the following assumption.
\begin{assum}\label{assum:SIA}
(Sparse Interference Assumption, SIA) For each unit $(r,c)$, there exists an index subset $\mathcal{N}_{rc}$ with cardinal $s_{rc}\ll RC$ such that for any $M_{\mathcal{N}_{-rc}}$ and $M'_{\mathcal{N}_{-rc}}$, the potential outcomes satisfy
$$Y_{rc}(M_{rc},M_{\mathcal{N}_{rc}},M_{\mathcal{N}_{-rc}})=Y_{rc}(M_{rc},M_{\mathcal{N}_{rc}},M'_{\mathcal{N}_{-rc}}).$$
\end{assum}

The smallest $\mathcal{N}_{rc}$ satisfying Assumption \ref{assum:SIA} is the interference neighbor set of interest.
We are particularly interested in detecting the elements in $\mathcal{N}_{rc}$ for each unit $(r,c)$ from the observational data, as this can provide insights into the mechanisms underlying treatment effects. Understanding these mechanisms can inform the development of more effective strategies.
The SIA covers many existing assumptions commonly used in the causal inference literature to address spatial interference, including the previously mentioned partial interference effect and local network interference \citep{RN115, RN33, RN126, RN4}. However, the SIA allows for variation in the interference structure across space, as long as the number of interference neighbors remains much smaller than $RC$.

In scenarios where prior knowledge suggests high-dimensional interference, one may refer to  \cite{leung2022causal} which assumes the interference effect decays as the distance between two units increases (referred to as approximate neighborhood interference, ANI). This assumption differs from our sparse interference assumption. Both assumptions have their respective advantages. While ANI potentially accommodates high-dimensional interference effects, our sparse assumption allows for distant interactions. Furthermore, \cite{leung2022causal} focuses on the asymptotic properties of the ATE estimator but does not provide non-asymptotic bounds or specific insights into the spatial structure of the effects, setting their work apart from ours. 
Additionally, when the interference decay rate is sufficiently fast, our proposed method remains applicable even when the underlying interference follows an ANI structure. We plan to explore this in further detail in subsequent revisions.

Under CA and SIA, the ITE can be expressed as
$$\text{ITE}_{rc}=\mathbb{E}\{Y_{rc}(M_{rc}=1,M_{\mathcal{N}_{rc}}=1)-Y_{rc}(M_{rc}=0,M_{\mathcal{N}_{rc}}=0)\}.$$
Let $X_{rc}$ denote the observed state variable for unit $(r,c)$. We now introduce the following classical assumption:

\begin{assum}
(Conditional Independence Assumption, CIA) The treatment assignment $M$ is independent of all the potential outcomes $Y(M=1)$ and  $Y(M=0)$ conditional on the the state variables $\{X_{rc}\}_{r,c}$.
\end{assum}

This assumption, also referred to as the unconfoundedness assumption, is fundamental for the identification of causal effects from observational data.  
Finally, under CA, CIA and SIA, the individual DE and IE can be expressed as 
\begin{eqnarray*}
\text{DE}_{rc}=\mathbb{E}_X\left\{\mathbb{E}(Y_{rc}|M_{rc}=1,M_{\mathcal{N}_{rc}}=1,X)-\mathbb{E}(Y_{rc}|M_{rc}=0,M_{\mathcal{N}_{rc}}=1,X)\right\}, \\ 
\text{IE}_{rc}=\mathbb{E}_X\left\{\mathbb{E}(Y_{rc}|M_{rc}=0,M_{\mathcal{N}_{rc}}=1,X)-\mathbb{E}(Y_{rc}|M_{rc}=0,M_{\mathcal{N}_{rc}}=0,X)\right\},
\end{eqnarray*}
with $\text{ITE}_{rc}$ and ATE derivable in a similar manner. For the remainder of this paper, we assume that the CA, CIA, and SIA hold.

\subsection{The proposed model}
\label{sec:model}

For $1\le r\le R$ and $1\le c\le C$, denote the $n$ independent observations in unit $(r,c)$ by $(Y_{i,rc}, X_{i,rc}, M_{i,rc})$, $i=1,\ldots,n$, where $Y_{i,rc} \in \mathbb{R}$ is the outcome, $X_{i,rc} \in \mathbb{R}^d$ is the state variable, $M_{i,rc} \in \left\{-1,1\right\}$ is the treatment variable. The underlying model is assumed to be
\begin{equation}\label{model:L+S}
    Y_{i,rc} = X_{i,rc}\beta_{rc} + M_{i,rc}L_{rc} + M_{i,-rc}S_{rc} + \varepsilon_{i,rc},\  i = 1, \dots,n,
\end{equation}
where $\beta \in \mathbb{R}^d$ is the effect of state variables, $L_{rc}\in \mathbb{R}$ is the DE at $(r, c)$, and $S_{rc} \in \mathbb{R}^{RC-1}$ represents the IE at $(r, c)$. The random errors $\{\varepsilon_{i,rc}\}_i$ have mean zero and variance $\sigma^2_{rc}$ and may exhibit correlation across $i$. 

Let $J_1(r,c)$ be the index set of nonzero elements in $S_{rc}$.  There is a one-to-one correspondence between $\mathcal{N}_{rc}$ and $J_1(r,c)$, and the SIA aligns with the classical sparse assumption in high-dimensional linear models. Within the potential outcome framework, it is straightforward to verify that $\text{DE}_{rc}=L_{rc}$,\ $\text{IE}_{rc}=\sum_{j\in J_1(r,c)} S_{rc,j}$ and
$$\text{ATE}=\frac{1}{RC}\sum_{r,c}\left\{L_{rc}+\sum_{j\in J_1(r, c)}S_{rc,j}\right\}.$$
Once the sets $\{J_1(r,c)\}_{r,c}$ are accurately identified, model \eqref{model:L+S} reduces to an ordinary linear regression model, enabling effective inference on the effects of interest.
Consequently, the primary objective of this paper is to detect locations of IE. Given that the estimates of IE and DE are correlated--both depending on treatment variables--it is crucial that the DE estimates converge at a rate faster than root-$n$ in order to efficiently detect IE. 

To achieve this, we propose incorporating spatial information, such as low-rank, smoothness, or clustered-uniform structures, to improve the accuracy of DE estimation. In this paper, we specifically focus on the generalized low-rank structure
in the sense that $L \in \mathcal{B}_q(v_q)$, where 
\begin{equation*}
    \mathcal{B}_q(v_q) = \left\{ A \in \mathbb{R}^{m_1\times m_2}: \sum_{i=1}^m \theta_i^q(A) \leq v_q , \quad m = \min\left\{m_1, m_2\right\}\right\},
\end{equation*}
with $\{\theta_i(A), i = 1, \dots, m\}$ denoting the singular values of $A$. 
The set $\mathcal{B}_q(v_q)$ provides flexibility in modeling various structures. When $q = 0$, $\mathcal{B}_0(v_0)$ corresponds to the class of matrices with at most $v_0$ nonzero singular values. As shown in \cite{chernozhukov2021inference}, generalized low-rank matrix regression encompasses a wide range of latent-variable models, capable of accommodating heterogeneous treatment effects. Furthermore, the main idea can be adapted to alternative assumptions regarding  the direct effect by adjusting the penalty term accordingly. For example, similar to \cite{luo2024policy}, one can also adopt the smoothing techniques to enhance the accuracy of DE estimation.

\section{Estimation and Detection Procedures}
\subsection{Estimation procedure}\label{sec:estimate}

We introduce the estimation procedure for model \eqref{model:L+S} in this subsection.
Let $Y_{rc} = (Y_{1,rc}, \dots, Y_{n,rc})^\top $, $X_{rc} = (X_{1,rc}, \dots, X_{n,rc})^\top $, $M_{rc} = (M_{1,rc}, \dots, M_{n,rc})^\top $ and $M_{-rc} = (M_{1,-rc}, \dots, M_{n,-rc})^\top $. 
The target function for the low-rank and sparse model is
{\small\begin{equation}\label{target fn}
    Q(\beta, L, S) = \frac{1}{n}\sum_{r=1}^R\sum_{c=1}^C\norm{Y - X_{rc}\beta_{rc} - M_{rc}L_{rc} - M_{-rc}S_{rc}}_F^2 + \lambda\norm{L}_* + \sum_{r=1}^R\sum_{c=1}^C\lambda_{rc}\norm{S_{rc}}_1,
\end{equation}}
where  $\|\cdot\|_F, \|\cdot\|_*$ and $\|\cdot\|_1$ are the Frobenius norm, nuclear norm and 1-norm, respectively.

We propose to use a two-layer profiling algorithm to solve the minimization of the target function \eqref{target fn}. Firstly, given $\beta^{(t)}$ and $S^{(t)}$, let $\tilde{Y}_{rc}^{(t)} = Y_{rc} - X_{rc}\beta_{rc}^{(t)} - M_{-rc}S_{rc}^{(t)}$, $r = 1, \dots, R; c = 1, \dots, C$. Define
\begin{equation}\label{Loss1}
    L^{(t+1)} = \arg\min_{L}\frac{1}{n}\sum_{r=1}^R\sum_{c=1}^C\norm{\tilde{Y}_{rc}^{(t)} - M_{rc}L_{rc}}_F^2 + \lambda\norm{L}_* \overset{\triangle}{=} \arg\min_L \mathcal{L}^{(t)}(L) + \lambda\norm{L}_*,
\end{equation}
then $L^{(t+1)}$ can be solved by a standard FISTA step proposed in \cite{RN165}, which combines a gradient descent method with a soft threshold on the nuclear norm of the estimator. Specifically, let $\tilde{L}^{(t)} = L^{(t)} + (r^{(t-1)} - 1)(L^{(t)} - L^{(t-1)})/r^{(t)}$, where $r^{(t)}$ is a scale parameter that is updated with the profiling procedure. The matrix soft threshold function is defined as, for any given matrix $D$ and threshold $\delta$,
\begin{equation}\label{soft fn}
    \mathrm{Soft}(D, \delta) = U_D^\top \mathrm{diag}\left\{\left(\theta_i(D) - \delta\right)_+\right\}V_D,
\end{equation}
where $D = U_D^\top \mathrm{diag}\left\{ \theta_i(D) \right\}V_D$ is the singular value decomposition of $D$. Next, $L^{(t+1)}$ is updated according to 
\begin{equation*}\label{update L}
    L^{(t+1)} = \mathrm{Soft}\left( \tilde{L}^{(t)} - \nabla \mathcal{L}^{(t)}\left(\tilde{L}^{(t)}\right)/\eta, \lambda/\eta \right),
\end{equation*}
where $\nabla \mathcal{L}^{(t)}\left(\tilde{L}^{(t)}\right)$ is the derivative of the loss function $\mathcal{L}^{(t)}$ on point $\tilde{L}^{(t)}$ with $\mathcal{L}^{(t)}$ defined in \eqref{Loss1}. Since the treatments are taken from $\left\{-1, 1\right\}$, thus the updating rate $eta$ is taken as $\eta = 2$ in our problem according to \cite{RN165}.
Next we denote $\bar{Y}_{rc}^{(t)} = Y_{rc} - M_{rc}L^{(t+1)} - M_{-rc}S_{rc}^{(t)}$ and estimate $\beta_{rc}^{(t+1)}$ by the ordinary least square (OLS) method,
$$\beta_{rc}^{(t+1)}=\arg\min_{\beta}\norm{\bar{Y}_{rc}^{(t)} - X_{rc}\beta_{rc}}_F^2/n.$$
Then let $\dot{Y}_{rc}^{(t)} = Y_{rc} - X_{rc}\beta_{rc}^{(t+1)} - M_{rc}L_{rc}^{(t+1)}$ and solve the lasso problem
\begin{equation*}\label{update S}
    S_{rc}^{(t+1)} = \arg\min_{S_{rc}}\frac{1}{n}\norm{\dot{Y}_{rc}^{(t)} - M_{-rc}S_{rc}}_F^2 + \lambda_{rc}\norm{S_{rc}}_1.
\end{equation*}
Finally, for a tolerance parameter $\tau$, we end the estimation procedure if 
\begin{equation*}
     \frac{\norm{\beta^{(t+1)} - \beta^{(t)}}_F}{\norm{\beta^{(t)}}_F} + \frac{\norm{L^{(t+1)} - L^{(t)}}_F}{\norm{L^{(t)}}_F} + \frac{\norm{S^{(t+1)} - S^{(t)}}_F}{\norm{S^{(t)}}_F} < \tau.
\end{equation*}

In the above procedure, the tuning parameter $\lambda$ for the low-rank penalty can be selected by the 5-fold cross validation. 
For the sparse tuning parameter $\lambda_{rc}$, we recommend to set $\lambda_{rc} = A\hat{\sigma}_{rc}\sqrt{\log(RC)/n}$ with $A > 2\sqrt{2}$, where $\hat{\sigma}_{rc}$ is some consistent estimator of $\sigma_{rc}$. We summarize the estimation algorithm in Algorithm \ref{alg:profiling}. We remark that the convergence of Algorithm \ref{alg:profiling} holds naturally due to the convergences of the FISTA, lasso, and profiling algorithms.

\begin{algorithm}
\caption{Two-layer Profiling Algorithm}\label{alg:profiling}
\begin{algorithmic}[1]
\State\textbf{Input}: observations $\{Y_{rc},X_{rc},M_{rc},M_{-rc}\}_{r,c}$; tuning parameters $\lambda, \{\lambda_{rc}\}_{r,c}$; updating rate $\eta=2$; scale parameters $r^{(-1)} = r^{(0)} = 1$; convergence tolerance $\tau$; initial estimators $\beta^{(0)}, S^{(0)}$, $L^{(0)}=L^{(-1)}$.
\State Initialize $t\leftarrow0$, $\delta^{(0)}\leftarrow1$.
\While{$\delta^{(t)}\ge\tau$}
    \For{$r=1,\ldots,R,\ c=1,\ldots,R$}
        \State Compute $\tilde{Y}_{rc}^{(t)} \leftarrow Y_{rc} - X_{rc}\beta_{rc}^{(t)} - M_{-rc}S_{rc}^{(t)}$ and $\tilde{L}^{(t)} \leftarrow L^{(t)} + \frac{r^{(t - 1) - 1}}{r^{(t)}}\left( L^{(t)} - L^{(t-1)} \right)$;
        \State With $\mathrm{Soft}(\cdot)$ and $\mathcal{L}^{(t)}$ defined in \eqref{soft fn} and \eqref{Loss1}, respectively, update 
        $$L^{(t+1)} \leftarrow\mathrm{Soft}\left( \tilde{L}^{(t)} - \Delta \mathcal{L}^{(t)}\left(\tilde{L}^{(t)}\right)/\eta, \lambda/\eta \right);$$
        \State Compute $\bar{Y}_{rc}^{(t)} \leftarrow Y_{rc} - M_{rc}L^{(t+1)} - M_{-rc}S_{rc}^{(t)}$ and update
        $$\beta_{rc}^{(t+1)}\leftarrow\arg\min_{\beta}\norm{\bar{Y}_{rc}^{(t)} - X_{rc}\beta_{rc}}_F^2/n;$$
        \State Compute $\dot{Y}_{rc}^{(t)}\leftarrow  Y_{rc} - X_{rc}\beta_{rc} - M_{rc}L_{rc}$ and update 
        $$S^{(t+1)}\leftarrow\arg\min_{S_{rc}}\frac{1}{n}\norm{\dot{Y}_{rc}^{(t)} - M_{-rc}S_{rc}}_F^2 + \lambda_{rc}\norm{S_{rc}}_1;$$
    \EndFor
    \State Update $r^{(t+1)} \leftarrow [1 + \{1 + 4\left( r^{(t)} \right)^2\}^{1/2}]/{2}$ and
    $$\delta^{(t+1)}\leftarrow\frac{\norm{\beta^{(t+1)} - \beta^{(t)}}_F}{\norm{\beta^{(t)}}_F} + \frac{\norm{L^{(t+1)} - L^{(t)}}_F}{\norm{L^{(t)}}_F} + \frac{\norm{S^{(t+1)} - S^{(t)}}_F}{\norm{S^{(t)}}_F};$$ 
    \State $t\leftarrow t+1$;
\EndWhile
\State \textbf{Output}: estimator $\hat{\beta}= \beta^{(t+1)}, \hat{L}= L^{(t+1)}$ and $\hat{S}= S^{(t+1)}$.
\end{algorithmic}  
\end{algorithm}

\subsection{Global test}\label{sec:test}

We consider the global testing problem for the interference tensor $S$ in this subsection. The hypothesis of interest is
\begin{equation}
\label{eq:global test}
    H_0:S = \boldsymbol{0}\quad \mathrm{v.s.} \quad H_1: S\neq \boldsymbol{0}.
\end{equation}
Let $\hat{S}$ be the estimator of $S$ under model \eqref{model:L+S}. 
We propose to construct the test statistic based on the infinity norm of the normalized estimator {\citep{zhang2017simultaneous, xue2020aos}.
Specifically, for a tensor $A\in\mathbb{R}^{p_1\times p_2\times p_3}$, define $\|A\|_\infty=\max\{|A_{ijk}|:i=1,\ldots,p_1;j=1,\ldots,p_2;k=1,\ldots,p_3\}$. Then the test statistic is  $T_n = \norm{\sqrt{n}\hat{S}}_\infty$. 
This test will reject the null hypothesis at certain significance level $\alpha$ if $T_n > c_B(\alpha)$ for some threshold $c_B(\alpha)$. The derivation of $c_B(\alpha)$ follows from a conditional bootstrap procedure. Specifically, let $\hat{\sigma}_{rc}$ be a consistent estimate of the noise levels $\sigma_{rc}$. We independently generate $e_{rc} = \left\{e_{irc} \right\}_{i=1}^n$ as a set of i.i.d. normal random variables from $\mathcal{N}(0, \hat{\sigma}^2_{rc})$, $r = 1, \dots, R; c = 1, \dots, C$, respectively. Then let
\begin{equation*}
    Y_{rc}^e = (I - \mathcal{P}_{X_{rc}})e_{rc},\ r = 1, \dots, R;\ c = 1, \dots, C,
\end{equation*}
where $\mathcal{P}_{X_{rc}} = X_{rc}\left(X_{rc}^\top X_{rc}\right)^{-1}X_{rc}^\top $ is the projection matrix of $X_{rc}$.
The rationale for constructing $Y_{rc}^e$ in this manner is as follows. First, taking advantage of the low-rank assumption, the estimation error of direct effect is negligible compared to $\widehat{\beta}_{rc}-\beta_{rc}$, making it unnecessary to recompute $L_{rc}$ in the bootstrap procedure. Second, it is worth noting that $\beta_{rc}$ is estimated using the OLS method. Given the convergence of profiling algorithms, it is equivalent to  utilize $(I - \mathcal{P}{X{rc}})e_{rc}$ for estimating $S_{rc}$ under the null hypothesis.
Let $\hat{S}^e_{rc}$ be the lasso estimator of the optimization problem $\norm{Y_{rc}^e - M_{-rc}S_{rc}}_F^2/n + \lambda_{rc}\norm{S_{rc}}_1$ and $T_n^e = \norm{\sqrt{n}\hat{S}^e}_{\infty}$. Then $c_B(\alpha)$ is calculated by
\begin{equation*}
    c_B(\alpha) = \inf\left\{ t\in\mathbb{R}: \mathbb{P}_e\left(T_n^e \leq t\right) \geq 1 - \alpha \right\},
\end{equation*}
where $\mathbb{P}_e\left(\cdot\right)$ denotes the probability measure with respect to $e = \left\{ e_{rc} \right\}_{r,c = 1}^{R,C}$ only. To approximate the critical value by the bootstrap, we generate $N$ sets of normal random variables  $e^{(1)}, \dots, e^{(N)}$ with each being a random copy of $e$. The corresponding estimates are denoted by $\hat{S}^e_1,\ldots, \hat{S}^e_N$. While keeping the state and treatment fixed, we calculate $N$ times of $T_n^e$, denoted as $\left\{ T_n^{eb}: b=1,\dots, N\right\}$ with $T_n^{eb}=\norm{\sqrt{n}\hat{S}^e_b}_{\infty}$. Then we approximate $c_B(\alpha)$ by the $100(1 - \alpha)$th sample percentile of $\left\{ T_n^{eb}: b=1,\dots, N\right\}$.

It remains to estimate the noise level $\sigma_{rc}$ of each unit,  $r = 1, \dots, R; c = 1, \dots, C$.
Similar to \cite{zhang2017simultaneous}, we adopt the scaled lasso which jointly estimates the regression coefficients and noise level in a linear model with the loss function
\begin{equation*}
    L_{\rho,rc}(\gamma_{rc})=\frac{1}{n}\norm{Y_{rc} - Z_{rc}\gamma_{rc}}_F^2 + \rho\norm{\gamma_{rc}}_1
\end{equation*}
where $Z_{rc} = \left( X_{rc}, M_{rc}, M_{-rc} \right)$ and $\gamma_{rc} = \left( \beta_{rc}, L_{rc}, S_{rc} \right)^\top $.


\subsection{Detection of interference}\label{sec:detection}

Note that the nonzero elements of $\hat{S}$ still include many false positive signals \citep{zhao2006irre}. We conduct a further detection procedure based on a bootstrap method.
Similar to the global test, the detection thresholds are generated from the $R\times C\times (RC-1)$-dimensional tensors $\{\hat{S}_b^e\}_b$, which determines the computational complexity. Hence the sparsity of $\hat{S}$ can not significantly reduce computational cost.
For ease of representation, we describe the detection procedure applied to the tensor $\hat{S}$, which is intrinsically the same as detecting within the nonzero elements of $\hat{S}$. 

Denote $U_n = \abs{\sqrt{n}\mathrm{vec}(\hat{S})}$ and $U_n^{(b)} = \abs{\sqrt{n}\mathrm{vec}(\hat{S}^e_{(b)})}$, where $\mathrm{vec}(\cdot)$ is the vectorization operator and $\hat{S}^e_{(b)}$ is the lasso estimator with respect to $e^{(b)}$ in the bootstrap procedure, $b = 1, \dots, N$. Then the stepdown algorithm \citep{Romano2005} can be applied to $U_n$, $U_n^{(1)}, \dots, U_n^{(N)}$ to consistently detect the interference. Specifically, let $\eta_1 = \{1, \dots, p\}$ at the first step where $p=RC(RC-1)$. Reject all hypotheses $H_{0,j}$ such that $U_{n,j} > c_B(\alpha;\eta_1)$, where the threshold $c_{B}(\alpha;\eta_l)$ is the empirical $1 - \alpha$ quantile of $\norm{U_n^{(b)}(\eta_l)}_{\infty}$ and $U_n^{(b)}(\eta_l)$ is the sub-vector which contains the elements with index in $\eta_l$. If no hypothesis is rejected, then stop. If some hypotheses are rejected, let $\eta_2$ be the set of indices for those hypotheses not being rejected at the first step. At step $l$, let $\eta_l \subset \eta_{l-1}$ be the subset of hypotheses that are not rejected at step $l-1$. Reject all hypotheses $H_{0,j}, j \in \eta_l$ satisfying that $U_j > c_{B}(\alpha;\eta_l)$. If no hypothesis is rejected, stop the algorithm. Denote the stopping point by $l^*$.  Then the  detected locations of IE is given by $\hat{J}_1=\eta_1 \setminus \eta_{l^*}$, where ``$\setminus$'' indicates the elements in $\eta_1$ but not in $\eta_{l^*}$.

An alternative method is the Binary and Re-Search (BiRS) algorithm proposed by \cite{zhang2023binary}. It involves a sequence of binary segmentation operations and dynamic tests to increase detection accuracy and reduce computation. The detection procedure is summarized as follows. At the first step, let $\eta_{1_1} = \left\{1, 2, \dots, \lfloor p/2 \rfloor\right\}$ and $\eta_{1_2} = \left\{\lfloor p/2 \rfloor + 1, \dots, p\right\}$ with $\lfloor \cdot\rfloor$ taking value of the  largest integer below. The critical value is $c_{B}(\alpha;\eta_1)$ with $\eta_1 = \eta_{1_1}\cup \eta_{1_2}$. Let $T_{1_k} = \norm{U_n(\eta_{1_k})}_{\infty}$ with $k = 1, 2$. If $T_{1_k} > c_{B}(\alpha;\eta_1)$, then $\eta_{1_k}$ possibly contains IE and needs further detection by binary segmentation; otherwise, there is no interference within region $\eta_{1_k}$. For the $l$-th binary search, there exist $n(l)$ possible signal segments $\eta_{l_1}, \dots, \eta_{l_{n(l)}}$,  $1 \leq n(l) \leq 2^l$. For $k = 1, \dots, n(l)$, when $T_{l_k} > c_{B}(\alpha;\eta_l)$, we take $\eta_{l_k}$ as the set containing interference for the next search. The segmentation stops if $\abs{\eta_{l_k}} = 1$. Denote the detected interference location set by $\hat{\eta}^{(0)}_{1}, \dots, \hat{\eta}^{(0)}_{K_0}$. Let $\hat{\eta}^{(0)} = \cup_{k=1}^{K_0}\hat{\eta}^{(0)}_k$. In the re-search procedure, substitute the variables in $\hat{\eta}^{(0)}$ in $U_n, U_n^{(b)}, b = 1, \dots, N$ with zeros and repeat the global test and binary search. Add the newly detected locations to $\hat{\eta}^{(0)}$. Repeat the re-search procedure until the global test accepts that there is no interference and denote the final set of detected locations of IE by $\hat{J}_1$.

Based on the detection results, we can establish a ``post-detection'' estimator of ATE using the OLS estimation in each unit if $s_{rc}\ll n$. Denote the detected locations of interference neighbors of unit $(r,c)$ by $\hat{J}_1(r,c)$ and let $\hat{s}_{rc}=\#\hat{J}_1(r,c)$. Note that the definition of ATE is
$$\text{ATE}=\frac{1}{RC}\sum_{r,c}\left\{L_{rc}+\sum_{j\in J_1(r, c)}S_{rcj}\right\}.$$
Let $\tilde{Z}_{rc} = \left(X_{rc}, M_{rc}, M_{-rc,\hat{J}_1(r,c)}\right)$. For each unit $(r,c)$, compute the $d+1+\hat{s}_{rc}$ dimensional OLS estimator  $\tilde{\gamma}_{rc} = \left( Z_{rc}^\top Z_{rc} \right)^{-1}Z_{rc}^\top Y_{rc}$. Then the post-detection estimator of ATE is given by
\begin{equation}\label{eq:ate est}
    \hat{\mathrm{ATE}} = \frac{1}{RC}\sum_{r=1}^R\sum_{c=1}^C\sum_{i = d+1}^{d+1+\hat{s}_{rc}} \tilde{\gamma}_{rci},
\end{equation}
where $\hat{s}_{rc} = \abs{\hat{J}_1(r, c)}$. Note that $s_{rc}\ll RC$, and the model dimension is substantially  reduced as long as $\hat{J}_1(r,c)$ covers ${J}_1(r,c)$ with a high probability.

\section{Theoretical Analysis}
\label{sec:thm}

In this section, we present the non-asymptotic upper bound of the estimation error for the treatment effect coefficient estimates proposed in Section \ref{sec:estimate}. Then we establish the theoretical guarantees of the global test under the null hypothesis, and analyze the power of the global test and the accuracy of detection.

\subsection{Estimation error bound}
We first impose the following assumptions on the random errors and treatment design.

\begin{assum}
\label{ass:random error}
    The random errors $\varepsilon_{rc}$, $r = 1, \dots, R$; $c = 1, \dots, C$ are independent sub-Gaussian random vectors with $\mathbb{E}\left(\varepsilon_{rc}\right) = \boldsymbol{0}$ and $\operatorname{Var}\left( \varepsilon_{rc} \right) = \sigma^2_{rc}I_n$. 
\end{assum}

\begin{assum}
\label{ass:ind-state-treat}
    For any fixed vector $\delta \neq 0$, $$2\norm{\mathcal{P}_{X_{rc}}\left(M_{rc},M_{-rc}\right)\delta}_2^2 < \norm{\left(M_{rc},M_{-rc}\right)\delta}_2^2, $$
    for $r = 1, \dots, R$; $c = 1, \dots, C$.
\end{assum}

\begin{assum}
    \label{ass:RE}
    For some integer $s_{rc}$ such that $1 \leq s_{rc} \leq RC$, recall that $J_1(r,c)$ is the set of nonzero elements of the coefficient $S_{rc}$ and $J_0(r, c)$ is the set of zero ones. Assume that the following condition holds: for any fixed $\delta \neq 0$ and $\norm{\delta_{J_0(r,c)}}_1 \leq 3\norm{\delta_{J_1(r,c)}}_1$,
    \begin{equation*}
        \kappa_{rc}(s_{rc}, 3) = \min_{\abs{J_1(r,c)} \leq s_{rc}} \frac{\norm{M_{-rc}\delta}_2}{\sqrt{n}\norm{\delta_{J_1(r,c)}}} > 0.
    \end{equation*}
\end{assum}

Assumption \ref{ass:random error} assumes the independent structure and the tails of the random errors in each unit, and Assumption \ref{ass:RE} is the restricted eigenvalue (RE) condition. Both are classical  assumptions in lasso problem \citep{buhlmann2011statistics,bickel_simultaneous_2009}. Assumption \ref{ass:ind-state-treat} imposes certain restrictions on the algebraic correlation between the state and treatment. In simpler terms, to control the estimation error, the correlation between the state and treatment can  not be too strong. This usually holds in experimental studies and is relatively easy to verify in observational studies. Based on these assumptions, we can derive the following results for the estimates of $\hat{L}$ and $\hat{S}$.

\begin{thm}
\label{thm:estimation error}
    Suppose that Assumption \ref{ass:random error} and  \ref{ass:ind-state-treat} hold, $L \in \mathcal{B}_q(v_q)$ and the regularization parameters $\lambda = 2\norm{\chi(M, \mathcal{P}, \varepsilon)}_{\mathrm{op}}/n$ and $\lambda_{rc} = A\sigma_{rc}\sqrt{\log (RC)/n}$ with $A > 2\sqrt{2}$, where $\norm{\cdot}_{\mathrm{op}}$ is the operator norm for matrices and the $(r, c)$-element of $\chi(M, \mathcal{P}, \varepsilon)$ is $\varepsilon_{rc}^{\top}(I - \mathcal{P}_{X_{rc}})M_{rc}$. Then when $S = \boldsymbol{0}$, there exist constants $K_1$ and $K_2$ such that
    \begin{equation*}
        \norm{\Delta(L)}_F = \norm{\hat{L} - L}_F \leq \max\left\{ K_1v_q^{1/2}\left(\frac{R+C}{n}\right)^{1/2 - q/4}, K_2\frac{(RC)^{1 - A^2/16}\log^{1/4}(RC)}{n^{1/2}}\right\},
    \end{equation*}
    holds with probability at least $1 - \exp\left\{ -\left(R + C\right) \right\}$.

    When $S \neq \boldsymbol{0}$ and $\norm{S_{rc}}_{0} = s_{rc}$, $r = 1,\dots, R$; $c = 1, \dots, C$, further assume that Assumption \ref{ass:RE} holds, then with probability at least $1 - \exp\left\{ -(R + C) \right\} - \exp\left\{ -(2\log RC - \log\log RC) \right\}$, we have
    \begin{equation*}
        \norm{\Delta(L, S)}_F = \sqrt{\norm{\hat{S} - S}_F^2 + \norm{\hat{L} - L}_F^2} \leq K_3\max_{r,c}\left\{ \sigma_{rc}\sqrt{s_{rc}RC\log(RC)/n} \right\},
    \end{equation*}
    where $K_3 = 32A/\min_{r, c}\kappa^2_{rc}(s_{rc}, 3)$.
\end{thm}

Theorem \ref{thm:estimation error} establishes the upper bounds of estimation errors of DE and IE estimates. When IE equals zero, the estimation error of DE estimate  is primarily influenced by the common low-rank regression estimation error bound. In this case, the error $\|\hat{L}_{rc} - L_{rc}\|_F$ is $o(1/\sqrt{n})$ when the matrix size $RC\gg n$. When IE is nonzero, the estimation error depends on the common lasso estimation error bound. These convergence rates guarantee that the parameters can be efficiently estimated even  when the treatment effects signals are weak, based on which we further analyze the size and power of our test method in the following subsections.

\subsection{Size analysis}

 Define $\tilde{G}_{irc} = \left(I_n - \mathcal{P}_{X_{irc}}\right)M_{i,-rc}$ and $\mathcal{G}_i = \left( \tilde{G}_{i11}^\top\varepsilon_{i11}, \dots, \tilde{G}_{iRC}^\top\varepsilon_{iRC} \right)^\top$, $i=1,\ldots,n;r = 1,\dots, R; c = 1, \dots, C$. The normalized sum of $\mathcal{G}_i$ is defined as
$\mathcal{S}_{n}^{\mathcal{G}} = n^{-1/2} \sum_{i=1}^n \mathcal{G}_i$.
Recall the global test problem is
\begin{equation*}
    H_0: S = 0 \quad \mathrm{v.s.} \quad H_1: S\neq 0,
\end{equation*}
and the detection procedures can be viewed as multiple tests for possible signal regions. Since we assume that the random error is sub-Gaussian and the treatments take values from $\left\{-1, 1\right\}$, then as discussed in \cite{xue2020aos}, it is easy to validate that there exists a constant $b_0 > 0$ such that $\min_j\mathbb{E}\left(\mathcal{S}_{nj}^{\mathcal{G}}\right)^2 \geq b_0 > 0$. Moreover, there exists a sequence of constants $B_n$ satisfy that, 
\begin{equation*}
    \max_{j}\sum_{i=1}^n\mathbb{E}\left(\abs{\mathcal{S}_{nj}^{\mathcal{G}}}^3\right)/n \leq B_n; \quad \max_{1\leq i \leq n}\max_j\mathbb{E}\left\{ \exp\left(\abs{\mathcal{S}_{nj}^{\mathcal{G}}}/B_n \right) \right\} \leq 2.
\end{equation*}
We further assume the following on the ratio between the number of units and the sample size. 

\begin{assum}
\label{ass:n-p-ratio}
There exists a constant $b_1$ such that $B_n^2\log^7\left( R^2C^2n \right)/n \leq b_1$.
\end{assum}

This suggests that the number of units can grow exponentially with the sample size, making the method suitable for ultra-high dimensional settings. Moreover, we assume the following.

\begin{assum}
    \label{ass:limitation exists}
    The limits $\lim_{n \rightarrow \infty} M_{-rc}^\top M_{-rc}/n = H_{rc} $, $r = 1, \dots, R$; $c = 1, \dots, C$ exist.
\end{assum}

\begin{assum}
    \label{ass:test-cov-treat-corre}
    For any $\delta$ satisfies that $\norm{\delta}_1 < K_3\max_{r,c}\left\{ \sigma_{rc}\sqrt{s_{rc}\log(RC)/n} \right\}$, we have 
    \begin{equation*}
    \norm{\mathcal{P}_{X_{rc}}M_{-rc}\delta}_{\infty} = o\left( n^{-1/2} \right).
    \end{equation*}
\end{assum}

Assumption \ref{ass:limitation exists} is commonly made for inference of lasso estimator, which implies that the treatments follow a stable strategy, which is easily satisfied in practice. Assumption \ref{ass:test-cov-treat-corre} imposes a weaker correlation structure between the treatments and state variables compared to Assumption \ref{ass:ind-state-treat}, in order to control the size of the global test. Then we have the following result.

\begin{thm}
    \label{thm:glo-size}
    Suppose that $L \in \mathcal{B}_q(v_q)$ and Assumption \ref{ass:random error} -- \ref{ass:test-cov-treat-corre} hold. The regularization parameters $\lambda = 2\norm{\chi(M, \mathcal{P}, \varepsilon)}_{\mathrm{op}}/n$ and $\lambda_{rc} = A\sigma_{rc}\sqrt{\log (RC)/n}$ with $A > 2\sqrt{2}$. Then under the null hypothesis, we have that 
    \begin{equation*}
    \begin{split}
        \abs{\mathbb{P}\left( T_n > c_B(\alpha) \right)  - \alpha} & \lesssim 2K_{4}B_n^2\left\{ \log^7\left(R^2C^2n\right)/n \right\}^{1/6} + K_{5}(\max \sigma_{rc})^{1/3}\left\{ \frac{\log^5 RC}{n^2} \right\}^{1/6}\\
        & + K_{6}\frac{\max\left\{ v_q^{1/2}n^{q/4}(R+C)^{1/2 - q/4}, (RC)^{1 - A^2/16}\log^{1/4}(RC) \right\}}{(b_0nRC)^{1/2}},
        \end{split}
    \end{equation*}
    with probability at least $1 - \exp\left\{ -(R + C) \right\} - \exp\left\{ -2\sqrt{\log(RC)} \right\}$. 
\end{thm}

This demonstrates the error bound between the reject probability of the proposed test and the size $\alpha$ under the null hypothesis. The first term represents the approximation error between the true distribution of $\mathcal{S}_{n}^{\mathcal{G}}$ and its Gaussian approximation, the second term refers to the approximation error of the bootstrap procedure, and the third term pertains to the bias term of the DE estimate in the profiling procedure. From Theorem \ref{thm:glo-size}, we can conclude that, under mild conditions, $\mathbb{P}\left( T_n > c_B(\alpha) \right) - \alpha \rightarrow 0$ as $n \rightarrow \infty$. 

\subsection{Power and detection accuracy}

To assess the efficiency of our test, we  make the following assumptions regarding the magnitude of the interference signal strength.

\begin{assum}
    \label{ass:signal strength}
    The strength of the true interference satisfies that
    \begin{equation*}
        \norm{S}_{\infty} \geq 2K_3\max_{r,c}\sigma_{rc}\sqrt{s_{rc}\log\left(R^2C^2n\right)/n}.
    \end{equation*}
\end{assum}

Denote the power of the global test by $\mathrm{Power}(T_n)$. Then we have the following theorem.

\begin{thm}
     \label{thm:power}
     Suppose that Assumption \ref{ass:random error} -- \ref{ass:signal strength} hold. The regularization parameters are $\lambda = 2\norm{\chi(M, \mathcal{P}, \varepsilon)}_{\mathrm{op}}/n$ and $\lambda_{rc} = A\sigma_{rc}\sqrt{\log (RC)/n}$ with $A > 2\sqrt{2}$. Then we have that
     \begin{equation*}
         \mathrm{Power}(T_n) \geq 1 - 2n^{-2} \rightarrow 1,
     \end{equation*}
     with probability at least $1 - \exp\left\{ -(R + C) \right\} - \exp\left\{ -(2\log RC - \log\log RC) \right\}$.
\end{thm}
 
Theorem \ref{thm:power} guarantees that the power of global test tends to 1 even if the effects are weak. To see this, note that the interference is sparse and $\max_{r,c} s_{rc} \lesssim \log(RC)$ holds generally. Then Assumption \ref{ass:signal strength} requires the signal strength to be approximately $O(\log(RC)/\sqrt{n})$, which is fairly weak. 

We next study the detection accuracy of the stepdown and BiRS algorithms. We introduce the Jaccard index to measure the similarity between two sets $I_1$ and $I_2$ as
\begin{equation*}
    \mathcal{J}\left(I_1, I_2\right) = \abs{I_1 \cap I_2}/\abs{I_1 \cup I_2}.
\end{equation*}
Let $J_1$ be the index set of nonzero elements in $\mathrm{vec}(S)$. Recall that $\hat{J}_1$ is the detected index set of nonzero elements in $\mathrm{vec}(S)$. We say that $J_1$ is consistently detected by $\hat{J}_1$ if there exists some $\zeta = o(s)$ such that 
\begin{equation*}
    \mathbb{P}\left\{ \mathcal{J}\left(J_1, \hat{J}_1\right) \geq 1 - \zeta \right\} \rightarrow 1
\end{equation*}
as $s \rightarrow \infty$. We impose some mild assumptions on the stepdown and BiRS algorithms to ensure consistent detection of the interference. 

\begin{assum}
   \label{ass:treatment design}
    Let $s = \sum_{r=1}^R\sum_{c=1}^C s_{rc}$: \\
    (a) For any $(r, c) \neq (j, k)$, as $n \rightarrow \infty$,
   \begin{equation*}
         \frac{1}{n}\abs{M_{rc}^\top M_{jk}} \leq \frac{1}{K_3\max_{r,c}\sigma_{rc}s_{rc}};
   \end{equation*}
    (b) For any $(r, c) \neq (j, k)$, as $n \rightarrow \infty$,
   \begin{equation*}
       \frac{1}{n}\abs{M_{rc}^\top M_{jk}} = o\left(\frac{1}{K_3\max_{r,c}\sigma_{rc}s_{rc}}\right);
   \end{equation*}
 \end{assum}

 \begin{assum}
   \label{ass:detection signal}
   The strength of the true interference satisfies that,
   \begin{equation*}
       \min\left\{\abs{S_{J_1}}\right\} \geq 2K_3\max_{r,c}\sigma_{rc}\sqrt{s_{rc}\log\left(R^2C^2n\right)/n}.
   \end{equation*}
\end{assum}

Assumption \ref{ass:treatment design} specifies the requirements on the correlation of treatments across regions to ensure the consistent detection of stepdown and BiRS algorithms under (a) and (b), respectively, which can guide the experimental design in practice.  Assumption \ref{ass:detection signal} mandates that the interference strength is uniformly larger than the lower bound in Assumption \ref{ass:signal strength}. Based on these assumptions, we present the following theorem.

\begin{thm}
 \label{thm:detection accuracy}
   Let $\hat{J}_1^{\mathrm{step}}$ be the estimated nonzero interference set by stepdown algorithm. Suppose that Assumptions \ref{ass:random error} -- \ref{ass:test-cov-treat-corre}, \ref{ass:treatment design}(a) and Assumption \ref{ass:detection signal} hold, then we have that 
  \begin{equation}
  \label{eq:step_prob}
  \begin{split}
      & \mathbb{P}\left\{\mathcal{J}\left(J_1, \hat{J}_1^{step}\right) \geq 1 - \frac{\zeta}{s} \right\}  \geq \left\{ 1 - \frac{8\log(RC + 1)}{n^2} \right\} \times \\ 
      &\hspace{0.5in} \left\{ 1 - \sum_{k=\zeta}^{RC}\binom{RC}{k}\left(\frac{1}{\sqrt{\pi RC \log RC}}\right)^k\left(1 - \frac{1}{\sqrt{\pi RC \log RC}}\right)^{RC - k}\right\}.
  \end{split}
  \end{equation}
  Let $\hat{J}_1^{\mathrm{BiRS}}$ be the estimated nonzero interference set by BiRS algorithm. Suppose that Assumptions \ref{ass:random error} -- \ref{ass:test-cov-treat-corre}, \ref{ass:treatment design}(b), \ref{ass:detection signal} hold, we have that 
  \begin{equation}
  \label{eq:BiRS_prob}
      \mathbb{P}\left\{\mathcal{J}\left(J_1, \hat{J}_1^{BiRS}\right) \geq 1 - \frac{\zeta}{s} \right\} \geq \left\{ 1 - \frac{8\log(RC + 1)}{n^2}\right\}\left\{ 1 - K_6\left(\frac{\alpha}{\beta}\right)^{\zeta} \right\}.
  \end{equation}
 where $\beta$ and $K_6$ are constants with $\beta > \alpha$.
\end{thm}

To appreciate these results, we remark that the term $1 - 8\log(RC + 1)/n^2$ represents the probability that the estimated interference sets cover the true ones by either the BiRS or stepdown algorithm. Additionally, the second term in the right hand side (r.h.s.) of \eqref{eq:step_prob} and \eqref{eq:BiRS_prob}, respectively, provides the lower bound on the probability that $\zeta$ zero elements are falsely detected by the stepdown or BiRS algorithm, respectively.  Recall that $s=\sum_{r=1}^{R}\sum_{c=1}^Cs_{rc}$, where $R,C$ tend to be large. When  $s\sqrt{\log(RC)/RC} \rightarrow \infty$ and $\zeta = O(\sqrt{RC/\log(RC)})$, it holds that $\mathbb{P}\left\{\mathcal{J}\left(J_1, \hat{J}_1^{step}\right)\right\}\rightarrow1$. When $s\rightarrow \infty$ and $\zeta = o(s)$, it holds that $\mathbb{P}\left\{\mathcal{J}\left(J_1, \hat{J}_1^{BiRS}\right)\right\}\rightarrow1$. We mention that, although the stepdown algorithm can accommodate stronger correlation of treatments across regions, it tends to produce more conservative detection results than the BiRS algorithm, as evidenced in numerical experiments of Section \ref{sec:simulation}. This is probably because that the BiRS algorithm operates in shorter regions for detection and hence is empirically more efficient, especially when the signals are sparse. We recommend that the choice between the two algorithms should be made according to the situation at hand.

\section{Simulations}
\label{sec:simulation}

In this section, we conduct comprehensive real data based simulations to illustrate the usefulness of the proposed method. The data considered are the China meteorological forcing dataset (CMFD) \citep{CMFD,Ma2019atmosphere,Ma2019water}. The CMFD is a gridded near-surface meteorological dataset with high spatial and temporal resolutions that was specifically developed for analyzing land surface processes in China. This dataset is created by merging remote sensing products, reanalysis datasets, and in-situ observation data from weather stations. It provides a continuous record starting from January 1979 and extending until December 2018, with a temporal resolution of three hours and a spatial resolution of $0.1^\circ$. 
The geographical coverage ranges from $70^\circ \mathrm{E}$ to $140^\circ \mathrm{E}$ and from $15^\circ \mathrm{N}$ to $55^\circ \mathrm{N}$.

For our model, we select the 2-meter air temperature, surface pressure, specific humidity and 10-meter wind speed as the state covariates. The temperature is scaled to the unit $^\circ C$, while the pressure is scaled to the unit $\mathrm{KPa}$. The sample size is $n = 100$, and the samples are randomly selected from daily data in 2018. 
Our analysis focuses on the center of Hebei province in China, where we divide the area of interest into $R\times C$ units, 
where $(R,C)=(12,12)$ and $(16,16)$ as two different configurations. 
For the area with dimensions of $12 \times 12$, the latitude spans from $37.85^\circ\mathrm{N}$ to $40.05^\circ\mathrm{N}$ with an interval of $0.2^\circ$, while the longitude spans from $114.85^\circ\mathrm{E}$ to $117.05^\circ\mathrm{E}$ with the same interval. In the case of the area with dimensions of $16 \times 16$, the latitude extends from $37.45^\circ\mathrm{N}$ to $40.45^\circ\mathrm{N}$ with an interval of $0.2^\circ$, and the longitude extends from $114.45^\circ\mathrm{E}$ to $117.45^\circ\mathrm{E}$ with the identical interval.
We consider two scenarios for the designs of treatments: 
\begin{itemize}
    \item[(a)]  Independent design: randomly and independently assign treatments in each unit with a probability of $1/2$. This generates $n$ treatment matrices $M_1, \dots, M_{n}$.
    \item[(b)] Correlated design: independently generate random vectors $\tilde{M}_i$, where $i = 1, \dots, n$, from a multivariate normal distribution $\mathcal{N}(0, \Sigma)$, where $\Sigma = \left(0.5^{\abs{j - k}}\right)_{j, k = 1}^{RC}$. These vectors are then transformed into matrices, the treatment matrices $M_i$ are generated by taking the sign of the corresponding vectors $\tilde{M}_i$. 
\end{itemize}

To generate the matrix $M_{i,-rc}$, we employ the permutation operator $\mathcal{A}$. For $M_{i,-rc}$, the application of $\mathcal{A}$ on $M_{i,-rc}$ orders the treatments of other locations based on the Euclidean distances to location $(r,c)$. For illustrative purposes, consider a matrix $M = \left(m_{jk}\right)_{j, k=1}^3$. The operation of $\mathcal{A}$ on $M_{-11}$ is defined as follows:
\begin{equation*}
\begin{split}
    & M_{-11} = \left( m_{12}, m_{13}, m_{21}, m_{22}, m_{23}, m_{31}, m_{32}, m_{33} \right)^\top ; \\
    & \mathcal{A}\left(M_{-11}\right) = \left( m_{12}, m_{21}, m_{22}, m_{13}, m_{31}, m_{23}, m_{32}, m_{33} \right)^\top .
\end{split}
\end{equation*}

For generating regression coefficients, the state coefficient $\beta$ is drawn from a standard normal distribution. The DE matrix $L$ is generated based on the outcome matrix, conditioned on $M_{irc}=0$. Specifically, let $F_{rc} = \sum_{i=1}^n X_{irc}\beta_{rc}/n$ for $r = 1, \dots, R$ and $c = 1, \dots, C$.  Let $U_F^\top \mathrm{diag}\left\{\theta_i(F)\right\}V_F$ be the SVD of $F$. Then, the DE matrix $L$ is generated as
\begin{equation*}
    L = 0.01\times U_F^\top \mathrm{diag}\left\{\theta_1, \theta_2, \theta_3, \theta_4,0,\ldots,0\right\}V_F.
\end{equation*}
The generated $L$ exhibits a low-rank structure with a rank of $4$, and its effect on the outcome $Y$ is approximately $1\%$. In terms of the interference matrix $S$, for each unit $(r, c)$, we randomly select $\eta_{rc} = \left\{0, 1, 2\right\}$ clusters of units that interfere with unit $(r, c)$. Given unit $(r,c)$, a cluster is defined as a set of units that have the same distance to unit $(r, c)$. Additionally, given a strength parameter $\delta \in \left\{0, 2, 4, 6, 8, 10\right\}\times 10^{-3}$, the signal of nonzero IE is generated as follows,
\begin{equation*}
    S_{rcj} \sim \delta\cdot TN(0, \infty, \abs{F_{rc}}, 1), \quad j \in J_1(r,c),
\end{equation*}
where $TN(a, b, \mu, \sigma)$ is truncated normal distribution with support $(a, b)$, mean $\mu$ and standard deviation $\sigma$. The tuning parameter $\lambda$ for low-rank penalty is selected by 5-fold cross validation and the sparse tuning parameter $\lambda_{rc}$ is set to be $2\sqrt{3}\hat{\sigma}_{rc}\sqrt{\log(RC)/n}$.

We begin by comparing the size of our testing method with that of the debiased lasso test (DL-test) under the null hypothesis ($\delta = 0$). To calculate the empirical sizes, we conduct 1000 Monte Carlo simulations. Table \ref{table:size} displays the results, showing that the proposed test effectively control the size under both independent and correlated treatment designs with $(R,C)=(12,12)$ and $(R,C)=(16,16)$, while the size of DL-test seems unstable in independent designs. 

\begin{table}
  \setlength{\abovecaptionskip}{0.cm}
  \caption{The empirical size of the proposed test and DL-test in independent and correlated scenarios.   \label{table:size}} 
  \setlength{\tabcolsep}{4mm}
  \centering
    \begin{tabular}{ccccc}
    \toprule
    & \multicolumn{2}{c}{$R \times C = 12 \times 12$} & \multicolumn{2}{c}{$R \times C = 16 \times 16$}  \\
    \cmidrule(r){2-3} \cmidrule(r){4-5} 
    & Proposed & DL-test & Proposed & DL-test  \\
    \midrule
    Independent & 0.0475  & 0.0725 & 0.0400 & 0.0325  \\
    Correlated & 0.0525 & 0.0400  & 0.0500  & 0.0525  \\
    \bottomrule
    \end{tabular}%
\end{table}

Next, we assess the detection accuracy by combining our testing method, DL-test, with the BiRS and stepdown algorithms (referred to as Proposed-BiRS, Proposed-Step, DL-BiRS, and DL-Step, respectively). We compare their empirical false discovery rates (FDRs) and true positive rates (TPRs). Additionally, we compare the popular Knockoff method under the low-rank and sparse effect model. The FDRs and TPRs of each method are calculated with varying values of $\delta$ and $(R,C)$, using 500 Monte Carlo simulations. The corresponding results are presented in Figure \ref{fig:independent} and \ref{fig:correlated}, respectively.
In the independent treatment design, Proposed-BiRS exhibits the highest TPRs and effectively controls the FDRs. DL-BiRS has lower TPRs than Proposed-BiRS, with a significant reduction in TPRs when $R,C$ increase. This is likely due to the fact that the debiased lasso method typically requires the design matrix to have a Gaussian distribution, and its performance is affected when a binary treatment design is used.
The TPRs and FDRs of the methods using the stepdown algorithm for detection are both lower than those using the BiRS algorithm, which supports the argument in Section \ref{sec:thm}. The Knockoff method demonstrates the lowest overall detection accuracy. This is possibly because it is difficult to generate suitable knockoffs for a design matrix that only contains $-1$ and $1$.
In the correlated treatment design, when $(R,C)=(12 ,12)$, the TPRs of Proposed-BiRS are slightly inflated. These results indicate that the BiRS algorithm is preferred for less correlated treatment designs, while the stepdown algorithm can be applied to control the FDRs under more correlated designs. 

\begin{figure}
    \centering
    \includegraphics[width = 10cm]{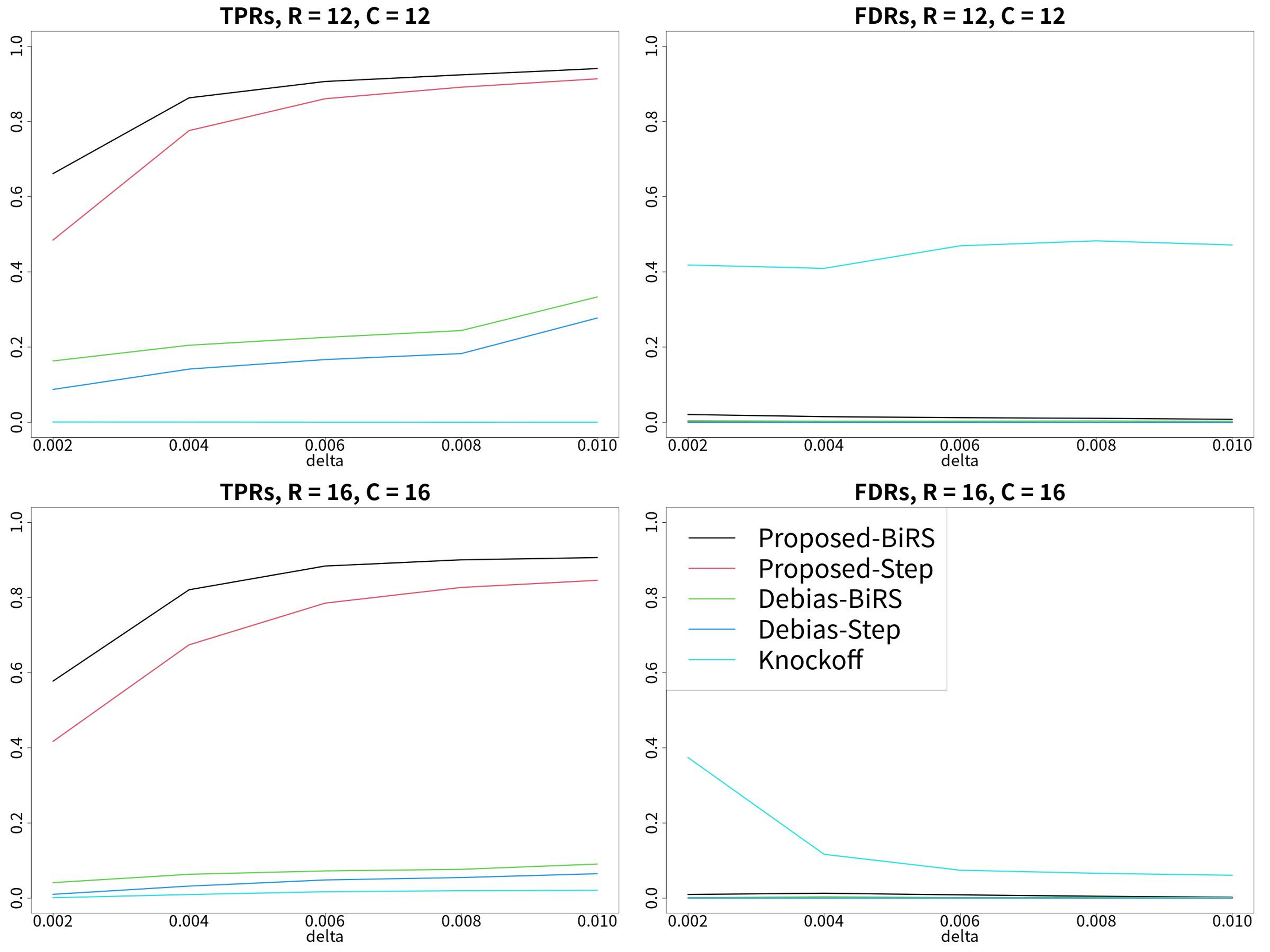}
    \caption{FDRs and TPRs of various methods under the independent treatment design for different values of $\delta$ ($\delta=0.002,0.004,0.006,0.008$, and $0.010$). The panels are organized into two rows corresponding to $(R,C)=(12,12)$ and $(R,C)=(16,16)$, respectively, while the columns display the TPRs and FDRs results, respectively.  }
    \label{fig:independent}
\end{figure}

\begin{figure}
    \centering
    \includegraphics[width = 10cm]{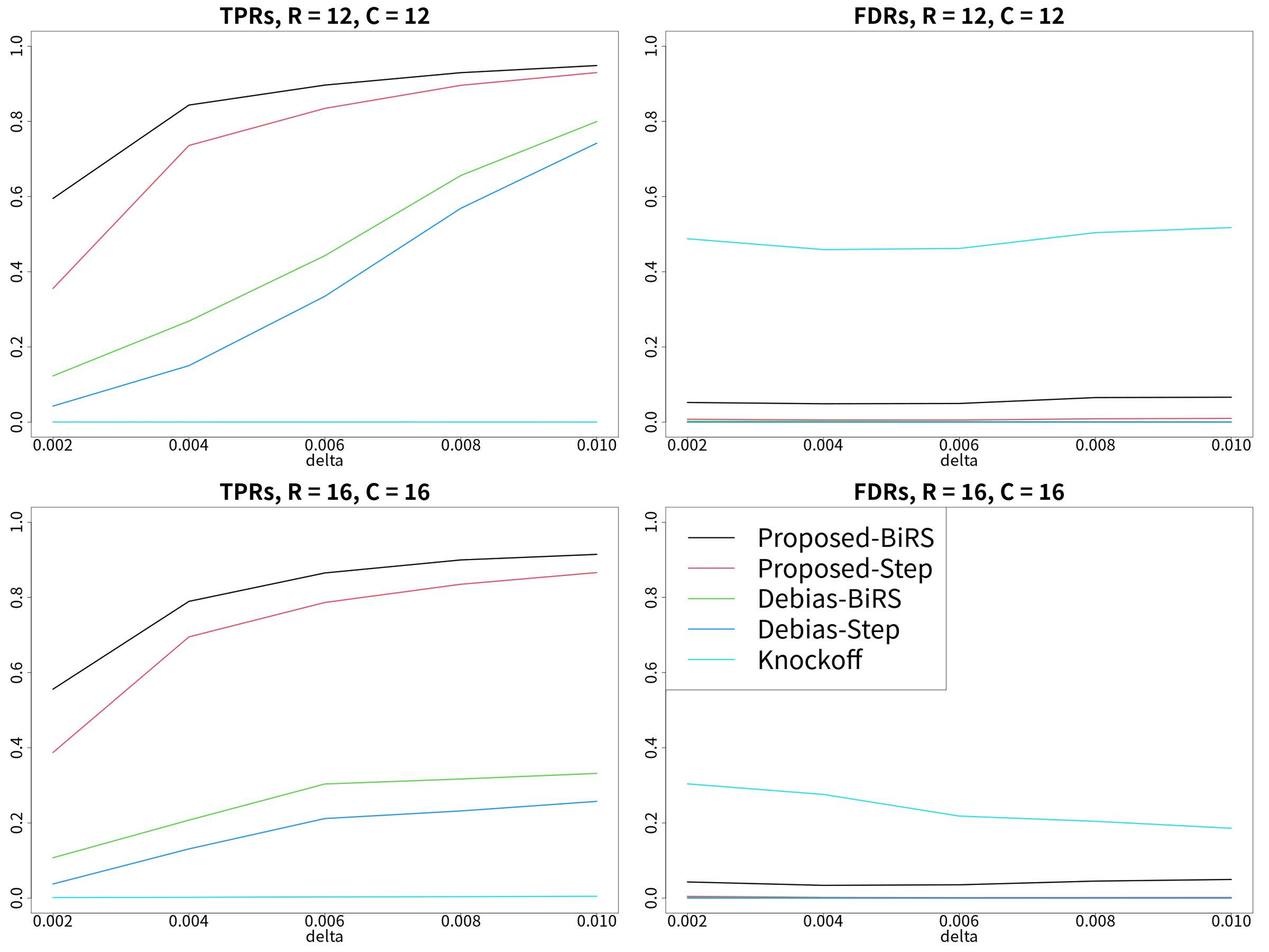}
    \caption{FDRs and TPRs of various methods under the correlated treatment design for different values of $\delta$ ($\delta=0.002,0.004,0.006,0.008$, and $0.010$). The panels are organized into two rows corresponding to $(R,C)=(12,12)$ and $(R,C)=(16,16)$, respectively, while the columns display the TPRs and FDRs results, respectively.}
    \label{fig:correlated}
\end{figure}

Lastly, we compare the estimation performance of the post-detection ATE estimate with the ATE estimate computed using the mean field approximation. The mean field approximation assumes that the interference neighbors of each unit are regions that share a common edge with it, as  discussed in \cite{shi2022multi} and \cite{hu2022average}. We calculate the empirical relative root mean square errors (RMSE) for both the post-detection ATE estimate and the mean field approximation ATE estimate across all previous settings, as depicted in Figure \ref{fig:RMSE-MS}. This indicates that the mean field approximation fails to achieve satisfactory performance when the interference structure is mis-specified in contrasted to the proposed method. 
We further investigate the relative RMSEs of the two estimates when the mean field assumption precisely holds. In this scenario, we define interference neighbors for each unit as those within a distance no greater than $\sqrt{2}$. The results are presented in Figure \ref{fig:RMSE-MF}.  Here the mean field estimate serves as an oracle in this setting, and its RMSE can be regarded as the benchmark for the proposed estimate. Figure \ref{fig:RMSE-MF} clearly illustrates that the post-detection ATEs closely approximate the oracle ATE as the signal strengths improve. This demonstrates the effectiveness of the proposed method in a range of scenarios, regardless of whether the common edge assumption holds or not.

\begin{figure}
    \centering
    \includegraphics[width = 10cm]{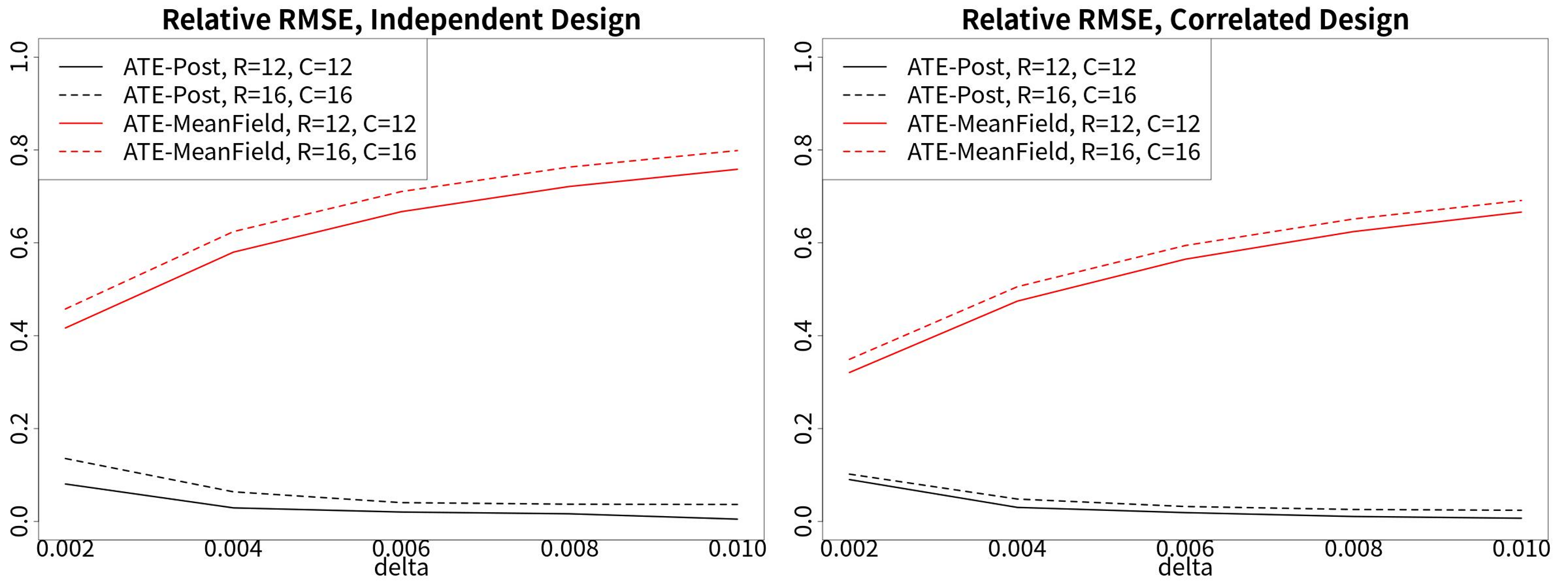}
    \caption{Relative RMSE of the post-detection ATE estimate (ATE-Post) and the mean field approximation ATE estimate (ATE-MeanField) under the independent design (left) and the correlated design (right) with $(R,C)=(12,12)$ and $(16,16)$ in mis-specified model. }
    \label{fig:RMSE-MS}
\end{figure}
 
\begin{figure}
    \centering
    \includegraphics[width = 10cm]{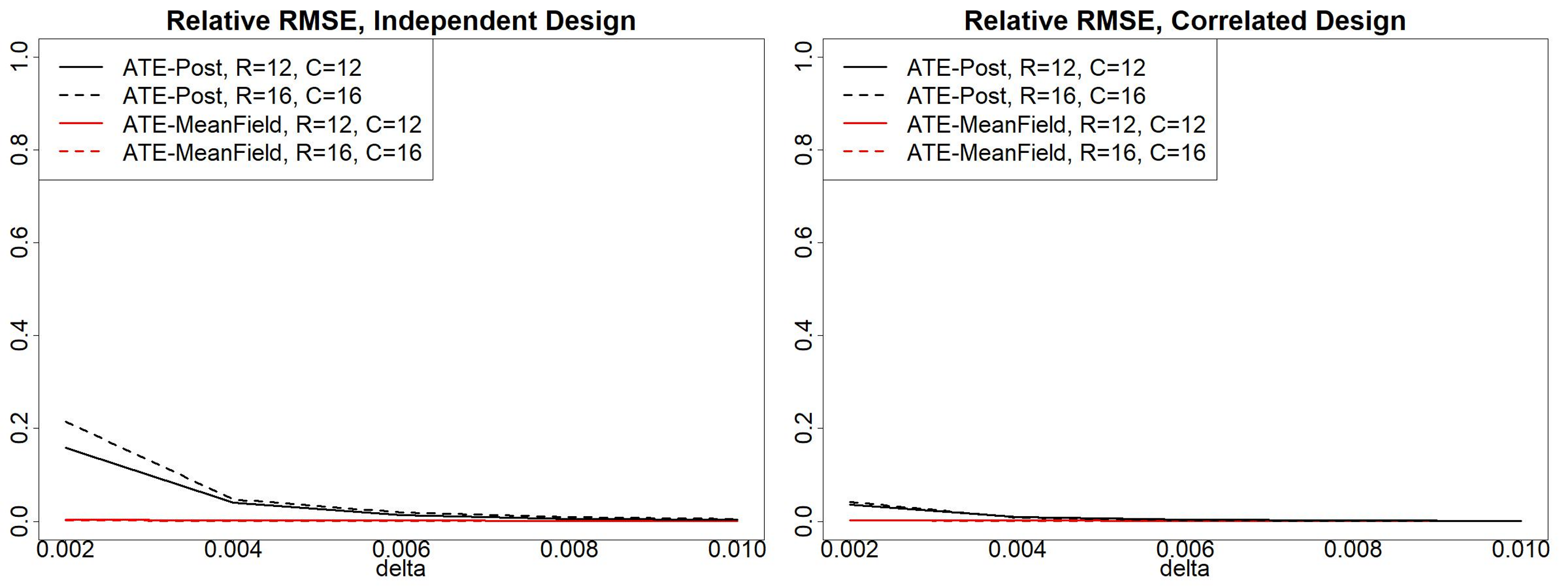}
    \caption{Relative RMSE of the post-detection ATE estimate (ATE-Post) and the mean field approximation ATE estimate (ATE-MeanField) under the independent design and the correlated design with $(R,C)=(12,12)$ and $(16,16)$ under mean field assumption. }
    \label{fig:RMSE-MF}
\end{figure}

\section{Applications}
\label{sec:applications}

In this section, we apply the proposed method to a typical scene of spatial causal inference: study the causal effect of meteorological elements on the concentration of PM10.

The data we utilize consist of daily PM10 concentration data for the year 2018. This dataset is derived from a series of comprehensive, high-resolution, and high-quality datasets of ground-level air pollutants for China, known as ChinaHighAirPollutants (CHAP) \citep{CHAPPM10,WEI2021106290}. The meteorological data for 2018 is obtained from the CMFD data introduced in Section \ref{sec:simulation}. Our primary focus is to investigate the causal relationship between wind speed and PM10 concentration. We consider the 2-meter air temperature and precipitation rate as the state variables and apply a one-day-lag to the PM10 concentrations to take the temporal autocorrelation into account according to \cite{Bartosz2017}. Wind speed is assigned a treatment value of $1$ if it exceeds $3.5m/s$ and $-1$ otherwise, as wind speeds below $3.5m/s$ are unlikely to generate dust. Our study area spans from $35.05^\circ \mathrm{N}$ to $40.05^\circ \mathrm{N}$ and $107.05^\circ \mathrm{E}$ to $117.05^\circ \mathrm{E}$. We use aligned grid cells in CHAP and CMFD datasets spaced $0.5$ degrees apart (approximately $50 km$), which results in a total of $11\times 21$ units. This region includes the Loess Plateau, a known source of PM10. 

\begin{figure}
    \centering
    \includegraphics[width = 11cm]{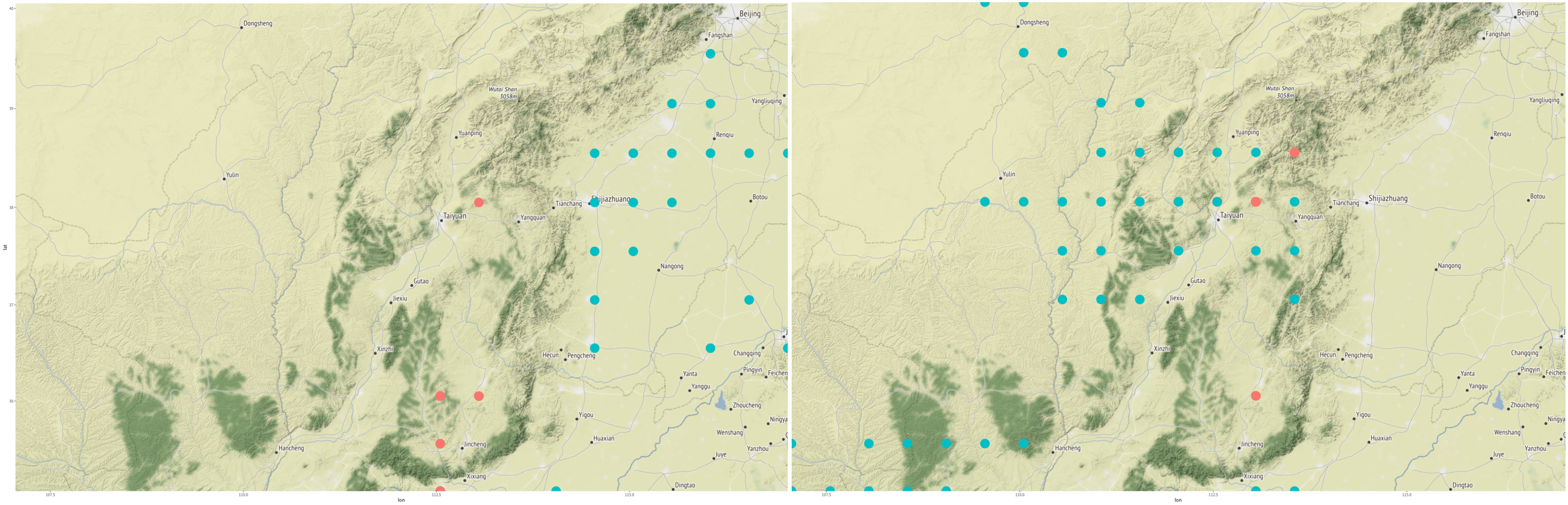}
    \caption{The interference structure of the causal effect of wind speed on PM10 concentration. In each panel, the treatments on regions plotted as red dots have an influence on the regions plotted as green dots.}
    \label{fig:CMFD}
\end{figure}

The interference structure we have identified is depicted in Figure \ref{fig:CMFD}. The pattern is strongly linked to the geographical terrain and defies the common assumption that interferences originate solely from neighboring units. The interference units we observed are all situated in the Taihang Mountains. The left panel of Figure \ref{fig:CMFD} demonstrates that wind speeds in the southern portion of the Taihang Mountains influence PM10 concentrations in the North China Plain. Conversely, the right panel of Figure \ref{fig:CMFD} reveals that wind speeds in the northern part of the Taihang Mountains have an effect on PM10 concentrations in neighboring and western units. It is worth noting that all interferences we observed are positive, suggesting that strong winds transport sand and dust from the Taihang Mountains to other areas, while the high Taihang Mountains obstruct the transport of PM10 between units.

\section{Concluding Remarks}
\label{sec:conclude}
In this work, we establish the framework for interference detection with theoretical guarantees. We introduce a low-rank and sparse model with a profiling algorithm, which is organically coupled with the high-dimensional signal detection method to identify the structure of spatial interference. We analyze the theoretical properties of the global test and detection procedure and demonstrate the effectiveness of our proposed method under mild assumptions. 

There are several important topics for future investigation. Firstly, we assume the noises $\{\varepsilon_{rc}\}_{r,c}$ are independent.  It would be intriguing to explore the situation that the noises are spatially correlated. 
The main challenge lies in deriving an upper bound for the operator norm of element-correlated matrices in the low-rank estimation. One potential solution is to assume that the spatial dependence is local. 
Secondly,  our focus in this work is on the parametric model. In the nonparametric setting, to address the curse of dimensionality, one may assume that the outcome of unit $(r,c)$ is related to the state and treatment variables through either a sparse partially linear model \citep{10.1214/07-AOS580,zhu2019high} 
, or a sparse single index model \citep{alquier2013sparse,naik2001single}. 
Lastly, we discuss post-detection ATE estimation in this work and the inference of post-detection ATE can be an interesting topic for future study.

\bibliographystyle{agsm}

\bibliography{Bibliography-MM-MC}
\end{document}